\documentclass[letterpaper,10pt]{article} 
\newcommand{\authorname}{Rahul Bhadani}

\newcommand{\booktitle}{Simultaneous Optimization of Channel Capacity}

\newcommand{\department}{Electrical and Computer Engineering\xspace}

\usepackage[colorinlistoftodos]{todonotes}

\usepackage{osameet3} %% use version 3 for proper copyright statement
\usepackage{rahulbhadani}
\usepackage[square,numbers]{natbib}
\graphicspath{ {figures/} }

\begin{document}

\title{\Large Programming the Kennedy Receiver for Capacity \\Maximization versus Minimizing One-shot Error Probability}

\author{Rahul Bhadani, Michael Grace, Ivan B. Djordjevic, Jonathan Sprinkle, Saikat Guha}
\address{Electrical \& Computer Engineering, College of Optical Sciences, \\The University of Arizona, Tucson, Arizona, USA}
\email{rahulbhadani@email.arizona.edu, michaelgrace@email.arizona.edu, ivan@email.arizona.edu, sprinkjm@email.arizona.edu, saikat@optics.arizona.edu}

%% Uncomment the following line to override copyright year from the default current year.
\copyrightyear{2020}

\begin{abstract}
We find the capacity attained by the Kennedy receiver for coherent-state BPSK when the symbol prior $p$ and pre-detection displacement $\beta$ are optimized. The optimal $\beta$ is different than what minimizes error probability for single-shot BPSK state discrimination.
\end{abstract}

\section{Optical communication capacity with coherent-state BPSK signaling}
\begin{figure}[htpb]
 \captionsetup[subfigure]{aboveskip=0pt,belowskip=1pt}
\centering
\begin{subfigure}{0.4\textwidth}
\centering
    \includegraphics[angle=0,origin=c,trim={0.0cm 0.0cm 0.0cm 0.0cm},clip,width=0.99\textwidth]{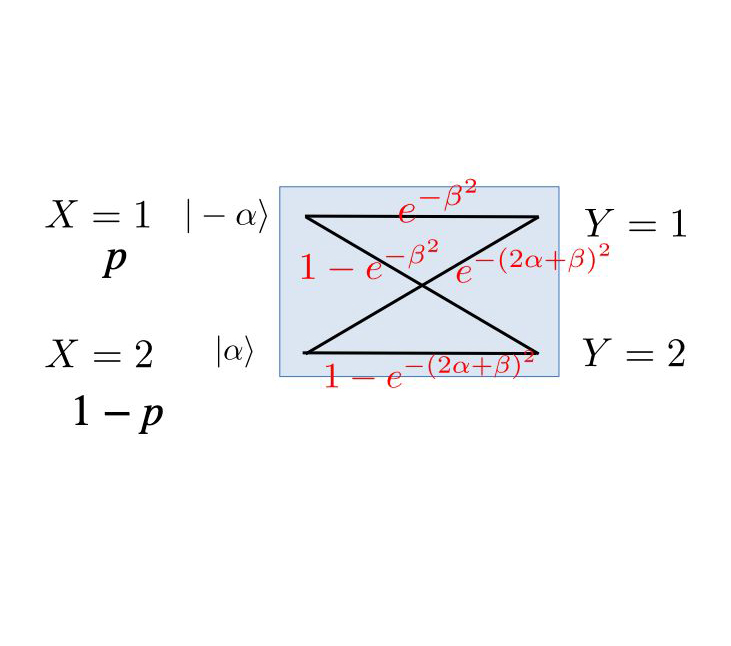}
 %   \caption{}
    \label{fig:txmatrix_optim_nulling}
\end{subfigure}
\begin{subfigure}{0.58\textwidth}
    \centering
    \includegraphics[angle=0,origin=c,trim={0.cm 7.0cm 0.0cm 6.0cm},clip,width=0.99\linewidth]{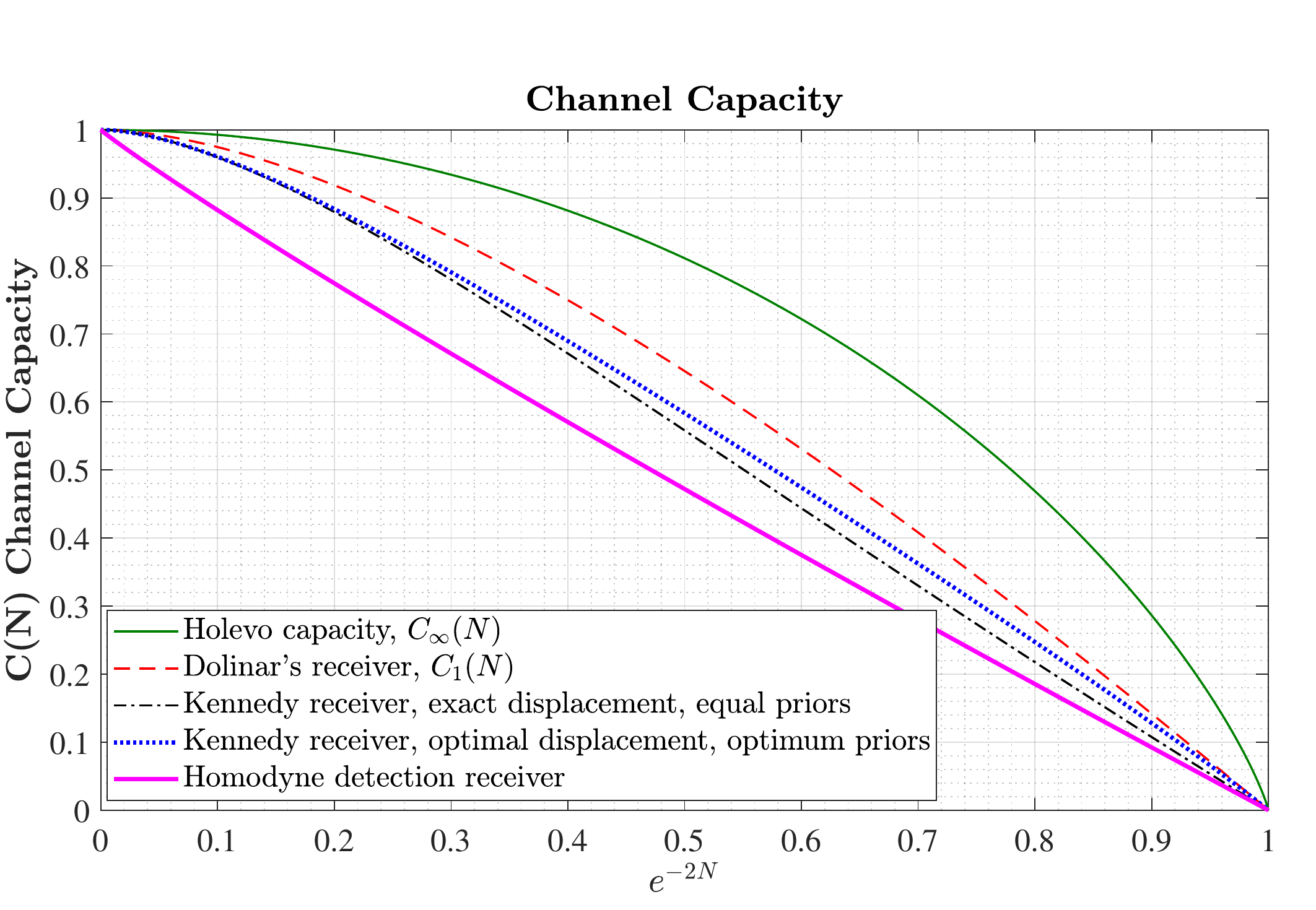}
%    \caption{}
    \label{fig:Channel_capacity_N10}
\end{subfigure}
\caption{[Left] Transition Probability Matrix for BPSK signaling and the Generalized Kennedy (GK) receiver. [Right] BPSK communication capacity attained by various receiver choices, in units of bits transmitted per BPSK symbol (bits per mode).}
    \label{fig:capacity}
\end{figure}

Let us consider coherent-state binary phase shift keying (BPSK) symbols $|\alpha\rangle, |-\alpha\rangle$, with priors $(p, 1-p)$ and mean photon number per mode, $|\alpha|^2 = N$, at the receiver. Sam Dolinar found the design of an optical receiver that uses a coherent-state local oscillator (LO), linear optics, switches, and quantum-noise-limited photon detection, which achieves the minimum average probability of error of discriminating the two states, $P_{e,{\text{min}}}(p) = \frac12[1-\sqrt{1-4p(1-p)e^{-4N}}]$~\cite{dolinar1976class}. For communicating with a BPSK alphabet, if a receiver must detect one BPSK symbol at a time, setting $p=1/2$, and using the Dolinar receiver achieves the maximum capacity, $C_1(N) = 1 - h_2(P_{e,{\rm min}}(1/2))$ bits per BPSK symbol, which is the capacity of a binary symmetric channel of crossover probability $P_{e,{\rm min}}(1/2)$. If the receiver is allowed to make joint (quantum, collective) measurements over long codeword blocks, not describable by any symbol by symbol measurement and post-processing, the maximum attainable capacity, $C_\infty(N) = h_2([1-e^{-2N}]/2)$ bits/symbol, is given by the Holevo limit~\cite{HSW}. In this article, we will restrict our attention to a specific (symbol-by-symbol) receiver design~\cite{Takeoka2008-jh} which is an extension of a receiver proposed by Robert Kennedy~\cite{kennedy1973near}. Hence our attained capacity will stay strictly below $C_1$. This generalized Kennedy (GK) receiver first applies a coherent displacement to shift the received BPSK symbols to $|\beta\rangle, |2\alpha + \beta \rangle$, and then detects the shifted states with a shot-noise limited photon detector. If the detector clicks, the receiver guesses the ``$|\alpha\rangle$" hypothesis, and if it generates no clicks, it guesses the ``$|-\alpha\rangle$" hypothesis. The pre-detection optical displacement can be realized by interfering the BPSK symbol on a beamsplitter of transmissivity $\kappa \approx 1$, with a strong coherent state LO $|\beta/\sqrt{1-\kappa}\rangle$. In Fig.~\ref{fig:capacity}, left, we show the $2$-by-$2$ transition probability matrix $p_{Y|X}(y|x)$ induced by the GK receiver. We calculate the capacity attained by the Kennedy receiver (by optimizing over the priors $p$ and the displacement $\beta$), i.e., $C_{\rm Kennedy}(N) = \textsf{max}_{p, \beta} I(X;Y)$. The plots in Fig.~\ref{fig:capacity} (right) show that the Capacity attained by the Kennedy receiver with optimized prior and displacement is higher than that achieved by Kennedy's original (exact displacement) receiver ($\beta = 0$), with $p$ optimized. The improvement is more pronounced when $N$ is small, the same regime where $C_\infty(N)/C_1(N) \to \infty$, of interest, e.g., for deep-space communications.

\section{Discussion of results: optimizing capacity versus minimizing the one-shot probability of error}

We now discuss our results in the context of important conceptual points regarding designing a receiver for maximizing (communication) rate, versus minimizing one-shot probability of error (e.g., in a target-detection radar).

\begin{figure}[htpb]
 \captionsetup[subfigure]{aboveskip=0pt,belowskip=-1pt}
\centering
\begin{subfigure}{0.46\textwidth}
\includegraphics[angle=0,origin=c,trim={0.cm 14.0cm 0.0cm 2.3cm},clip,width=1.0\linewidth]{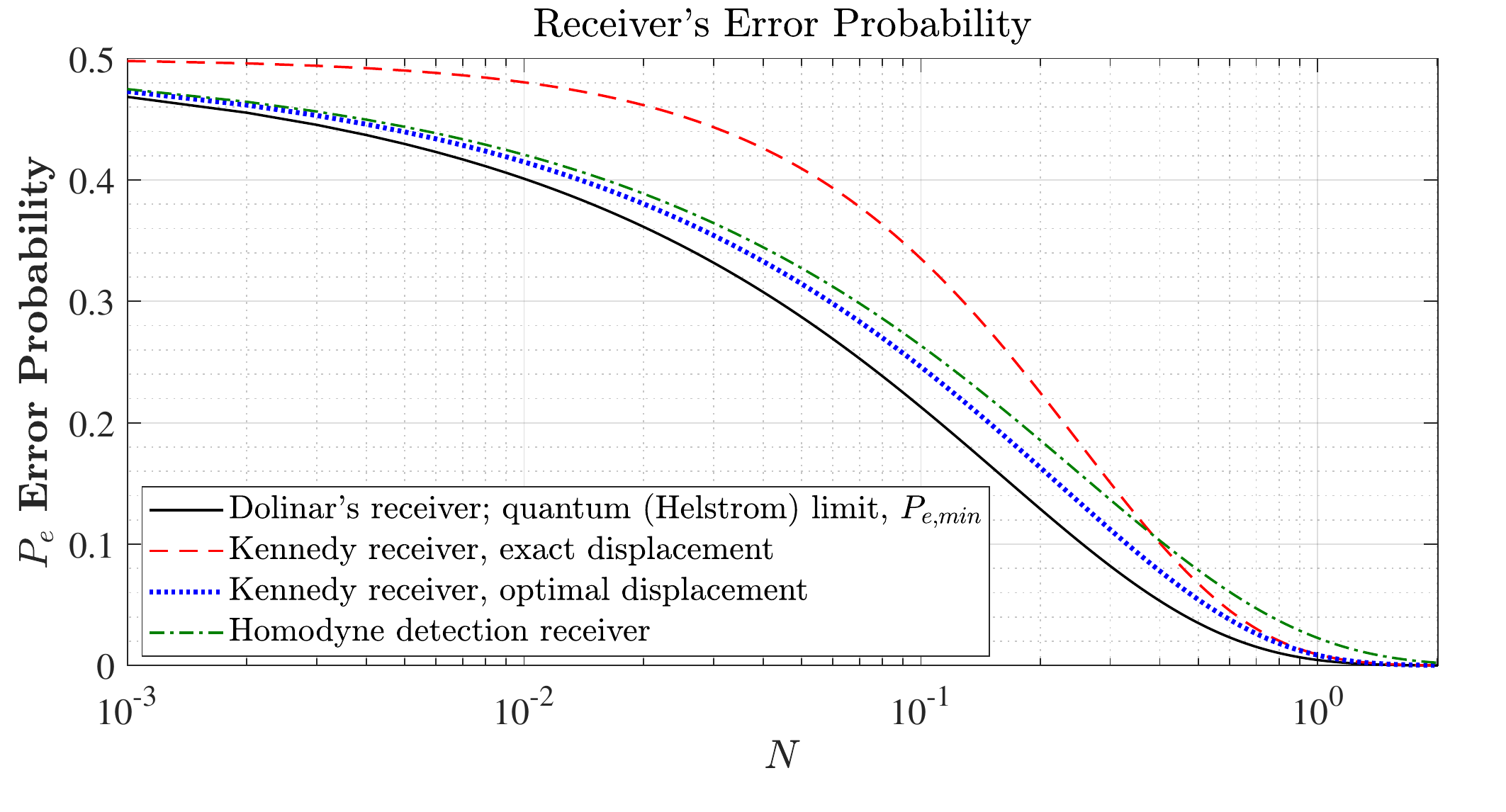}
\caption{}
\label{fig:Error_probabilityAndBeta_N10_1}
\end{subfigure}
\begin{subfigure}{0.48\textwidth}
\centering
\includegraphics[angle=0,origin=c,trim={0.cm 2.0cm 0.0cm 14.0cm},clip,width=1.0\linewidth]{Error_probabilityAndBeta_N10.pdf}
\caption{}
    \label{fig:Error_probabilityAndBeta_N10_2}
\end{subfigure}
    \caption{Figure~\ref{fig:Error_probabilityAndBeta_N10_1} shows average error probability for one-shot BPSK state discrimination, assuming equal priors. The quantum limit is given by the Helstrom limit, attained by Dolinar's receiver, which also attains capacity $C_1$, the optimal within all receivers that detect each BPSK symbol one by one.  Figure~\ref{fig:Error_probabilityAndBeta_N10_2} compares the $\beta$, optimal displacement for the Kennedy receiver, for: (a) optimizing capacity (red dashed), and (b) minimizing error probability assuming equal priors (blue dotted).}
    \label{fig:Error_probabilityAndBeta_N10}
\end{figure}

{\textbf{(a) Error probabilities for one-shot state discrimination.}} Figure~\ref{fig:Error_probabilityAndBeta_N10_1} shows average error probability for one-shot BPSK state discrimination, assuming equal priors, $p = 1/2$. The quantum (Helstrom) limit is attained by the Dolinar receiver. The Kennedy receiver's performance is shown, with optimized $\beta$. In Fig.~\ref{fig:Error_probabilityAndBeta_N10}, we see that the optimal $\beta$ is quite different than what minimizes error probability for single-shot BPSK state discrimination. Next, we note that Homodyne receiver's probability of error performance is strictly worse than the Kennedy receiver with optimized $\beta$, whereas there is a crossover between Homodyne and Kennedy receivers' performance when $\beta = 0$ is set. For channel capacity (see Fig.~\ref{fig:capacity}), both for $\beta = 0$ and $\beta$ optimized for capacity, homodyne receiver does inferior to the Kennedy receiver. This shows that an optical receiver may need to be ``programmed" differently based on the information-processing task at hand and that optimizing a receiver to minimize symbol error probability may not result in a capacity-maximizing setting for that same receiver, and vice versa.

{\textbf{(b) Finite blocklength communications rate.}} Bondurant generalized Kennedy's receiver to the $M$-ary phase-shift keying (PSK), later generalized to any $M$-ary constellation. Experiments have been conducted as well~\cite{Chen2012-fw, Becerra2015-uh}. This generalized receiver, known as sequential waveform nulling (SWN), does not meet the quantum limit of minimum error state discrimination. But it achieves the optimal error exponent, which is a factor of four higher than that achievable with ideal heterodyne detection~\cite{Nair2014-dr}. The symbol demodulation error probability $P_{e,{\rm Het}} \sim e^{-\xi N}$, $P_{e,{\rm SWN}} \sim e^{-4 \xi N}$, and $P_{e,{\rm opt}} \sim e^{-4 \xi N}$, as $N \to \infty$. Here $N$ is the average photon-number of the states in the constellation. In the high $N$ regime, there is not much capacity improvement by using SWN (over ideal heterodyne). But SWN's superior error exponent for symbol demodulation translates to a higher reliability function, i.e., a higher number of bits being transmissible over $n$ channel uses for a given target codeword error probability \cite{Tan2015-iz}. This could translate to a higher data volume being transmissible over a dynamic optical link that is only available for a short time duration. In ongoing work, we are optimizing displacement and nulling-order for $M$-ary constellations to maximize the finite-length rate, by optimizing the channel dispersion attained by the SWN receiver~\cite{polyanskiy2010channel}.

\section{Acknowledgment}
This was an {\em advanced problem} in the University of Arizona course OPTI 595B ``Information in a Photon", Spring $2019$. The research for Rahul Bhadani was supported by the National Science Foundation, award $1521617$.

%s\bibliographystyle{abbrvnat}

\bibliography{references2}

\end{document}